\documentclass[fleqn,usenatbib]{mnras}

\usepackage{newtxtext,newtxmath}

\usepackage[T1]{fontenc}
\usepackage{ae,aecompl}

\usepackage{graphicx}	
\graphicspath{{imgs/}{../imgs/}}
\usepackage{amsmath}	
\usepackage{mathtools}
\usepackage{listings}
\usepackage{xcolor}
\usepackage{array}
\usepackage[scr=rsfso,calscaled=.96]{mathalpha}
\usepackage{subfiles}
\usepackage{wasysym}

\defcitealias{DB98}{DB98}

\newcommand{\cmmnt}[1]{}

\title[The velocity distribution of white dwarfs in Gaia]{The velocity distribution of white dwarfs in Gaia EDR3}

\author[D. Mikkola, P. J. McMillan, D. Hobbs, J. Wimarsson]{
Daniel Mikkola\thanks{E-mail: mikkola@astro.lu.se }$^{1}$,
Paul J. McMillan$^{1}$, 
David Hobbs$^{1}$, 
John Wimarsson$^{2}$
\\
$^{1}$Lund Observatory, Lund University, Department of Astronomy and Theoretical Physics, Box 43, SE-22100, Lund, Sweden\\
$^{2}$Space Research \& Planetary Sciences, Physics Institute, University of Bern, Gesellschaftsstrasse 6, 3012 Bern, Switzerland}

\date{Accepted XXX. Received YYY; in original form ZZZ}

\pubyear{2019}

\begin{document}
\label{firstpage}
\pagerange{\pageref{firstpage}--\pageref{lastpage}}
\maketitle

\begin{abstract}

Using a penalised maximum likelihood we estimate, for the first time, the velocity distribution of white dwarfs in the Solar neighbourhood. Our sample consists of 129 675 white dwarfs within 500 pc in \textit{Gaia} Early Data Release 3
The white dwarf velocity distributions reveal a similar structure to the rest of the Solar neighbourhood stars, reflecting that white dwarfs are subjected to the same dynamical processes. In the velocity distribution for three magnitude-binned subsamples we however find a novel structure at $(U, V) = (7, -19)$ km s$^{-1}$ in fainter samples, potentially related to the \textit{Coma Berenices} stream. We also see a double-peaked feature in $U$-$W$ at $U \approx -30$ km s$^{-1}$ and in $V$-$W$ at $V \approx -20$ km s$^{-1}$ for fainter samples.
We determine the velocity distribution and velocity moments as a function of absolute magnitude for two samples based on the bifurcation identified in \textit{Gaia} Data Release 2 in the colour-magnitude diagram. The brighter, redder sequence has a larger velocity dispersion than the fainter, bluer sequence across all magnitudes. It is hard to reconcile this kinematic difference with a bifurcation caused purely by atmospheric composition, while it fits neatly with a significant age difference between the two sequences.
Our results provide novel insights into the kinematic properties of white dwarfs and demonstrate the power of analytical techniques that work for the large fraction of stars that do not have measured radial velocities in the current era of large-scale astrometric surveys.

\end{abstract}

\begin{keywords}
methods: statistical - methods: data analysis - Galaxy: structure - Galaxy: Solar neighbourhood - stars: kinematics and dynamics - Galaxy: kinematics and dynamics.
\end{keywords}



\section{Introduction}\label{sec:intro}

The present day structure and the history of the Galaxy are encoded not just in the positions of its stars but also in their kinematics. It is well established that the present velocity distribution in the Solar neighbourhood has a great deal of structure in it (e.g., \citealt{katz}) for which there are multiple possible causes. Suggested origins for overdensities include dissolving open clusters, resonances from large-scale density waves such as the Galactic bar and spiral arms, accreted populations from galaxy mergers, and phase mixing from nearby satellite galaxies (e.g., \citealt{antoja2012, kushniruk}). Understanding this substructure is a part of understanding the dynamical history of the Milky Way.

The phase-space distribution of stars within the Milky Way has been studied extensively over the last decades (See \citealt{katz} and references therein) to reveal this complicated structure, especially since the \textit{Hipparcos} mission \citep{hipparcos} and more so with its successor \textit{Gaia}'s \citep{gaia} recent second and third data release (Henceforth DR2 and EDR3 respectively, \citealt{dr2, edr3}).

The astrometry of \textit{Gaia} provides proper motions and positions for $\sim$1.5 billion sources with great precision, which is an enormous leap forward from its predecessor which observed $\sim$120 000 sources. The \textit{Gaia} data have only been available for a few years but the potential for kinematic study has already been demonstrated. For example,  \cite{oorts} accurately measured the Oort constants $A$ and $B$ as well as for the first time the non-axisymmetric constants $C$ and $K$. \cmmnt{In \cite{coma_b} it was shown that the moving group \textit{Coma Berenices} is limited to negative Galactic latitudes and likely has not undergone phase mixing in the Galactic potential. }The kinematic structure of the Solar neighbourhood has been studied in unprecedented detail to reveal many new and old structures (e.g, \citealt{kushniruk, katz}) as well as arches \cite{antojanature}. Beyond velocity space, the Solar neighbourhood has also been explored in orbit space (e.g., \citealt{trick1, trick2, trick3}) showing ridges which can manifest themselves as streams and structures in velocity space. Understanding the kinematic substructure of the Galaxy will require exploration in both velocity and orbit space, which becomes far more accessible due to the wealth of data provided by missions such as \textit{Gaia}.

Even though EDR3 provides accurate astrometry and photometry for a great number of sources, it does not contain full 3D phase-space information for all of them as some lack measured radial velocity. This means that for most individual stars only the position and proper motions are available as in the \textit{Hipparcos} catalogue. In fact, the number of sources with radial velocities in \textit{Gaia} EDR3, $\sim$7.2 million, is dwarfed by the number of sources with at least position and proper motions, $\sim$1.5 billion. This means that the radial velocity sample contains only $\sim$0.5\% of sources with the full astrometric solution.

This limitation can be circumnavigated by studying properties of an entire sample rather than the individual stars. This was demonstrated in two seminal papers, \cite{DB98} (hereafter  \citetalias{DB98}) and the follow-up paper \cite{wd98} for \textit{Hipparcos}. \citetalias{DB98} calculated the mean motion and velocity dispersions for populations taken from a sample of 11 865 single main sequence stars; and \cite{wd98} estimated the velocity distribution $f(\pmb{v})$ for a similar sample of stars. Both of these papers use the approximation that the velocity distribution is consistent for the full sample and is spread across the full sky.

As previously mentioned there are $\sim$1.5 billion sources in EDR3. This sample includes stars in \textit{Gaia}'s $G$-band magnitude system as faint as $G\approx 21$ which means that it contains samples of stars which have been previously unavailable for kinematic studies. This includes the white dwarfs (WDs) which, following DR2, revealed that the colour-magnitude diagram (CMD) of WDs has more structure than previously thought and displays a clear bifurcation as well as a crystallization branch (\citealt{dr2HR}, \citealt{tremblay19}).

Since then there have been a few different explanations put forward and we will briefly summarise a few here. Shortly after DR2 released \cite{el-badry18} used a cross-match of the 100 pc \textit{Gaia} WD sample with the Montreal White Dwarf Database (MWDD, \citealt{dufour}) to show that part of the bifurcation could be explained with an initial-final mass ratio (IFMR) that produces a bimodal WD mass distribution with peaks at both $\sim0.6$M$_\odot$ and $\sim0.8$M$_\odot$ if the age distribution in multimodal. The different mass WDs would populate different cooling tracks and produce the bifurcation. This would not produce a bifurcation in mono-age clusters however as massive WDs would cool before young WDs appear on the CMD.

Around the same time \cite{kilic18} used a similar sample and showed that atmospheric composition explains the bifurcation well for 0.6 M$_\odot$ WDs. However, they also conclude that the WD mass distribution is indeed bimodal. They suggest that this bimodality can be explained at least partly through the merger of WD binaries. More recently \cite{kilic2020} revisited the 100 pc WD sample and conducted a spectroscopic follow-up survey, thereby being able to constrain atmospheric composition reliably. They found that the mass distribution of WDs That have H lines as the primary feature of their spectrum (DA WDs), has a sharp peak at 0.59 M$_\odot$ with a broad shoulder, best fit with a secondary Gaussian at 0.76 M$_\odot$, again demonstrating the existence of a bimodal mass distribution. They test a WD model including mergers and find that it cannot produce a good fit to the observed mass distribution. They also investigate the transverse velocities of WDs, since merger products should appear as massive WDs with larger velocities than those formed from single main-sequence stars. However, the lack of young, massive WDs with large velocities coupled with the model predictions leads them to conclude that mergers are unable to explain the bimodal mass distribution of WDs. Instead, it is shown in \cite{tremblay19}, \cite{bergeron}, and \cite{kilic2020} that the effects of crystallisation is able to create the over-abundance of massive WDs in the 0.7-0.9 M$_\odot$ range. In this scenario, the massive WDs which should already have reached the bottom of the WD sequence are subjected to cooling delays (for a detailed explanation of these effects, see e. g. \citealt{bauer2020, blouin2021}).

The bifurcation in the WD CMD is well described by atmospheric differences and the bimodal WD mass distribution can arise due to core crystallisation and its related effects. We explore a new direction to probe the WD bifurcated sequences using their kinematics which could provide additional insight into the bifurcation problem. While kinematics could be affected in complex ways by the processes that produce binary merger systems, the second, fainter, sequence of the CMD should have the same velocity dispersion as the brighter sequence if it is caused by mergers or atmospheric composition. Previously \cite{rowell} performed a kinematic study of WDs in DR2 using the method of \citetalias{DB98} to determine the mean velocity and velocity dispersion of the WDs. However, as they only split the WD CMD in M$_G$ their results reflect a mixture of the two sequences and we expand upon this analysis by splitting the WDs across the visible bifurcation and computing the full velocity distributions in addition to velocity moments.

The paper is organised as follows: In Section \ref{sec:theory} we briefly present the techniques of \citetalias{DB98} and \cite{wd98} used to determine the moments and velocity distribution.  The WD samples that we have used and how they are selected is presented in Section \ref{sec:sample}. Then in Section \ref{sec:moments_results} we present the moments for the bimodal sequences and in Section \ref{sec:fv_results} the velocity distribution of nearby WDs is presented for the first time. The implication and significance of our results is discussed in Section \ref{sec:disc} and our conclusion are in Section \ref{sec:conc}.

\section{Theory}\label{sec:theory}

\subsection{Moments of phase space}\label{subsec:moments_theory}
In Galactic dynamics, we decompose the space velocity into velocity towards the Galactic centre, $U$, in the direction of rotation, $V$, and, north of the Galactic plane, $W$. Galactic observations however, use a combination of the line-of-sight velocity, $v_r$, and the combined on-sky velocity, $\pmb{p}$. With the release of Gaia we have access to a large sample of stars for which is measured positions ($\ell$, $b$), parallax ($\varpi$), and proper motions ($\mu_{\ell\ast}$, $\mu_b$). A subset of these will also have measurements of radial velocity $v_r$ and with these properties combined one can determine a star's space velocity $\pmb{v}_i$. Without the full three-dimensional velocity vector we can only know a star's tangential velocity $\pmb{p}_i$. Despite this it is still possible to determine the moments of phase space, the mean velocity components and the velocity dispersions, and we do this by making use of a deprojection technique from  \citetalias{DB98}.

To correct for Galactic rotation we use equation (1) of \citetalias{DB98} with values for Oort's constants taken from \cite{oorts} ($A= 15.3$ km s$^{-1}$ kpc$^{-1}$, and $B= -11.9$ km s$^{-1}$ kpc$^{-1}$). We express the tangential velocity in three dimensional Galactic coordinates as
\begin{equation}
	\pmb{p} = \frac{1}{\varpi}
	\begin{pmatrix*}
	-\sin\ell\ \mu_{\ell\ast} - \cos\ell\ \sin b\ \mu_b \\
	 \cos\ell\ \mu_{\ell\ast} - \sin\ell\ \sin b\ \mu_b \\
	 \cos b\ \mu_b
	\end{pmatrix*},
\end{equation}
and it is related to the space velocity through the projection
\begin{equation}\label{eq:proj}
	\pmb{p} = \mathbf{A}\pmb{v}.
\end{equation}
The transformation matrix, $\mathbf{A}$, is defined by
\begin{equation}
	\mathbf{A} \equiv \mathbf{I} - \pmb{\hat{r}} \cdot \pmb{\hat{r}}^T,
\end{equation}
where $\mathbf{I}$ is a 3x3 identity matrix and $ \pmb{\hat{r}}$ is the unit vector to the star, given by
\begin{equation}
	 \pmb{\hat{r}} =
	 \begin{pmatrix*}
	 	\cos b\ \cos \ell \\
	 	\cos b\ \sin \ell \\
	 	\sin b \\
	 \end{pmatrix*}.
\end{equation}
The symmetric matrix $\mathbf{A}$ is a projection operator and is thus singular and non-invertible which means that equation \eqref{eq:proj} cannot be inverted. This comes as no surprise; we cannot determine the space velocity of a star $\pmb{v}$ with its tangential velocity alone.
\subsubsection{Mean velocities}\label{subsubsec:v_av}
If the on-sky positions $\pmb{\hat{r}}$ of a sample of stars are uncorrelated with their velocities $\pmb{v}$, then so is the projection matrix. Then by taking the average of equation\eqref{eq:proj}
\begin{equation}
	\langle\pmb{p}\rangle = \langle\mathbf{A}\pmb{v}\rangle = \langle\mathbf{A}\rangle\langle\pmb{v}\rangle,
\end{equation}
the matrix $ \langle\mathbf{A}\rangle$ \textit{can} be inverted, which means that the average space velocity can be determined from the average tangential velocity
\begin{equation}
	\langle\pmb{v}\rangle = \langle\mathbf{A}\rangle^{-1}\langle\pmb{p}\rangle.
\end{equation}
The assumption of uncorrelated positions and velocities holds under the approximation that the velocity distribution is constant over the volume in question.

\subsubsection{Velocity dispersions}\label{subsubsec:disp}
We can now calculate the motion relative to the mean, known as the peculiar velocity
\begin{equation}\label{eq:peculiar}
	\pmb{p}' \equiv \pmb{p} - \mathbf{A}\langle\pmb{v}\rangle, \qquad 	\pmb{v}' \equiv \pmb{v} - \langle\pmb{v}\rangle,
\end{equation}
which allows us to determine the following 3x3 matrix
\begin{equation}
	\mathbf{B} = \langle\pmb{p}'{\pmb{p}'}^T\rangle = \frac{1}{N}\sum_{i=1}^{N}\pmb{p}'_i{\pmb{p}'}^T_i.
\end{equation}
In a similar manner to how we could reconstruct $\langle\pmb{v}\rangle$ from $\langle\pmb{p}\rangle$ we can estimate the dispersion tensor $\mathbf{D}$ from the matrix $\mathbf{B}$.

Combining equation \eqref{eq:peculiar} with equation \eqref{eq:proj} and writing it in component form, using Einstein summation convention gives
\begin{equation}
	p'_k = \mathbf{A}_{km}v'_m.
\end{equation}
This allows us to specify the different components of $\mathbf{B}$ as
\begin{equation}
	B_{kl} =  \langle p'_k p'_l\rangle =   \langle A_{km}v'_m A_{ln}v'_n\rangle = \langle A_{km}A_{ln}\rangle \langle v'_mv'_n\rangle
\end{equation}
where in the final step we again use the assumption of independence. We get the elements of the dispersion matrix as $D_{mn} = \langle v'_mv'_n\rangle$ and can simply write
\begin{equation}\label{eq:tmat}
	B_{kl} =\langle A_{km}A_{ln}\rangle D_{mn}.
\end{equation}
This can be inverted to find the elements of the dispersion tensor that correspond to the second order moment $(D_{11}, D_{22}, D_{33}) = (\sigma_U^2, \sigma_V^2, \sigma_W^2)$.

This approach is very similar to that of \citetalias{DB98} with the only difference being that here the sample is not assumed to be perfectly isotropic. This method has seen use in e. g. \cite{rowell} for a similar sample of stars to ours.

\subsection{Inferring velocity distributions}\label{subsec:fv_theory}
Just as we were able to determine the mean velocity and velocity dispersions from the tangential projection of the space velocities, we can infer the velocity distribution of a sample of stars without known radial velocities under the approximation that distribution is consistent for the whole sample and is spread across the sky. This method of finding $f(\pmb{v})$ was demonstrated already by \cite{wd98} for \textit{Hipparcos} stars. Since we use the same method we will outline only key details here. Consider now the probability distribution of tangential velocities in a given direction $\pmb{\hat{r}}$ as $\rho(\pmb{q}|\pmb{\hat{r}})$ and express it in terms of the full velocity distribution with the integral
\begin{equation}
	\rho(\pmb{q}|\pmb{\hat{r}}) = \int \mathrm{d}v_r f(\pmb{v}) =  \int \mathrm{d}v_r f(\pmb{p} + v_r\pmb{\hat{r}}).
\end{equation}
An estimate of the true distribution $f_0(\pmb{v})$ is the result of maximizing the log-likelihood for a given model of it, $f(\pmb{v})$,
\begin{equation}
		\mathscr{L}(f) = N^{-1}\sum_{k=1}^N \ln P(\pmb{q}_k | \pmb{\hat{r}}_k, f),
\end{equation}
where
\begin{equation}\label{eq:prob}
	P(\pmb{q}_k | \pmb{\hat{r}}_k, f) = \int \mathrm{d}v_r f(\pmb{p}_k + v_r\pmb{\hat{r}}_k)
\end{equation}
is the probability for a star $k$, in direction $ \pmb{\hat{r}}_k$ and with velocity drawn from $f(\pmb{v})$, to be observed with tangential velocity $\pmb{p}_k$. In principle this log likelihood could be maximised with a distribution function that has a series of delta functions, one for each star. We therefore introduce a penalty function to enforce smoothness. This function is given a weight $\alpha$ that acts as a smoothing parameter.
\begin{equation}
	\mathscr{Q}_\alpha(f) = \mathscr{L}(f) - \frac{1}{2}\alpha\mathscr{S}(f),
\end{equation}
where $\mathscr{S}(f)$ is the penalty function and a measure of the smoothness of $f(\pmb{v})$. There are two constraints to the function which maximizes $\mathscr{Q}_\alpha(f)$: It must be non-negative,
\begin{equation}\label{eq:nonneg}
	f(\pmb{v}) \geq 0,
\end{equation}
and its must be unity,
\begin{equation}\label{eq:unity}
	\mathscr{N}(f) \equiv \int \mathrm{d}^3\pmb{v}f(\pmb{v}) = 1.
\end{equation}
As is shown in \cite{wd98} we can meet these conditions in rather elegant ways. The condition of eq. \eqref{eq:unity} is met by maximizing
\begin{equation}
	\tilde{\mathscr{Q}}_\alpha(f) \equiv \mathscr{Q}_\alpha(f) - \mathscr{N}(f),
\end{equation}
instead of $\mathscr{Q}_\alpha(f)$ and the condition of eq. \eqref{eq:nonneg} is met by defining
\begin{equation}
	f(\pmb{v}) \equiv e^{\phi(\pmb{v})},
\end{equation}
with $\tilde{\mathscr{Q}}_\alpha$ now a function of $\phi(\pmb{v})$.

The numerical approach to the problem is as follows: view $\phi(\pmb{v})$ on a three-dimensional grid that has $L_U \times L_V \times L_W$ cells and widths of $h_U \times h_V \times h_W$. With each cell denoted by $\pmb{l}$ comprising three integers which label the cell, we then write
\begin{equation}\label{eq:numphi}
	\phi(\pmb{v}) = \sum_{\pmb{l}} \phi_{\pmb{l}} W_{\pmb{l}}(\pmb{v})
\end{equation}
where $W_{\pmb{l}}(\pmb{v})$ is the window function
\begin{equation}
	W_{\pmb{l}}(\pmb{v}) =
	\begin{dcases}
		\frac{1}{h_Uh_Vh_W}, & \mathrm{if}\ \forall i |v_i - l_ih_i - v_i^0| \leq \frac{1}{2}h_i,  v_i \in \{U, V, W\} \\
		0, & \mathrm{otherwise,}
	\end{dcases}
\end{equation}
where $\pmb{v_i^0}$ is the velocity at the centre of the first cell ($\pmb{l} = (0, 0, 0)$), meaning all cells positions are relative to it in $\pmb{v}$-space. This criterion simply means that the window function is zero if the velocity components do not fit into the grid cell in question. Combining equations \eqref{eq:numphi} and \eqref{eq:prob} we have
\begin{equation}
	P(\pmb{q}_k | \pmb{\hat{r}}_k, f) = \sum_{\pmb{l}} e^{\phi_{\pmb{l}}} K(k|\pmb{l}),
\end{equation}
where $K(k|\pmb{l})$ is simply the length of the segment of the line $\pmb{v} = \pmb{p}_k + v_r\pmb{\hat{r}}_k$ that lies in cell $\pmb{l}$, divided by $(h_Uh_Vh_W)$. The penalty function is approximated as:
\begin{equation}
	\mathscr{S}(f) \simeq \int\mathrm{d}^3\pmb{v}\left(\sum_{\pmb{n}} \phi_{\pmb{n}}\Xi_{\pmb{n}\pmb{l}}\right)^2,
\end{equation}
where
\begin{equation}
	\Xi_{\pmb{n}\pmb{l}} = \sum_{i=x,y,z}\frac{\tilde{\sigma}^2_i}{h_i^2}(-2\delta_{\pmb{n},\pmb{l}} + \delta_{\pmb{n},\pmb{l} + \pmb{\hat{e}}_i} + \delta_{\pmb{n},\pmb{l} - \pmb{\hat{e}}_i}).
\end{equation}
Here, $ \pmb{\hat{e}}_i$ denotes the unit vector in the $i$th direction. Putting it all together gives us the numerical approximation to the function we seek to maximize:
\begin{multline}\label{eq:functional}
	\tilde{\mathscr{Q}}_\alpha(\pmb{\phi}) = N^{-1}\sum_{k} \ln \left[\sum_{\pmb{l}}e^{\phi_{\pmb{l}}}K(k|\pmb{l})\right] - \sum_{\pmb{l}}e^{\phi_{\pmb{l}}} \\ - \frac{1}{2}\alpha h_xh_yh_z\sum_{\pmb{l}}\left(\sum_{\pmb{n}} \phi_{\pmb{n}}\Xi_{\pmb{n}\pmb{l}}\right)^2.
\end{multline}
\subsubsection{Maximizing the likelihood}\label{subsubsec:maximizing}
In order to maximize the value of eq. \eqref{eq:functional} we make use of the conjugate gradient method (e. g. \citealt{numericalrecipes}). The value of the smoothing parameter $\alpha$ in Eq. \eqref{eq:functional} determines the balance between goodness of fit and smoothness. In order to determine an appropriate value we use the \textit{Gaia} RVS sample. We create two subsamples by randomly selecting 130 000/30 000 sources (to match our {\tt all\_500} and {\tt red\_500}/{\tt blue\_500} sample sizes) and maximize the likelihood with a range of $\alpha$ values to determine the optimal value given different number of sources. Since the actual velocity distribution can be determined for the RVS sample we can determine an appropriate $\alpha$ by visual inspection. We find that the values $\alpha = 10^{-11}$ and $\alpha = 3\cdot10^{-11}$ appropriately reconstructs the velocity distribution and use it for our WD and Red/Blue samples respectively.

We reduce the number of operations required considerably by using an approach with an increasing grid size. This method uses the initial guess of $\pmb{\phi}$ on a crude Cartesian grid. The solution, $\hat{\pmb{\phi}}$, that maximizes $	\tilde{\mathscr{Q}}_\alpha$ is then interpolated on a finer grid and used as the initial guess for a new maximization on that finer grid. We allow for up to six different grids, each grid being twice as big in each dimension. This means the initial grid, $\pmb{n}_\mathrm{initial}$ will be of shape $\pmb{n}_\mathrm{final} / 2^k$ where $k$ is the number of grid steps and is chosen such that no dimension in $\pmb{n}_\mathrm{initial}$ is less than 10.\cmmnt{If any dimension in $\pmb{n}_\mathrm{initial}$ is not even the steps are rounded. This means that for a requested final grid of $102^3$ the grids will be $13^3$, $26^3$, $52^3$, and finally $104^3$. For our velocity distributions we use a grid of $\pmb{n} = [100, 100, 72]$ ranging} Our velocity distribution is found, following two stages of refinement, on a grid of $\pmb{n} = [100, 100, 72]$ over the range $U \in [-150, 150]$ km s$^{-1}$, $V \in [-150, 50]$ km s$^{-1}$, and $W \in [-80, 60]$ km s$^{-1}$.

Our setup is broadly consistent with the one used by \cite{wd98} with a few differences. Our algorithm is built without the original rejection criterion that any star's $K(k|\pmb{l})$ must pass through 96 cells. This means our distributions might take into account stars which lie outside of the grid. Instead of discarding these stars, we simply omit the outermost layers of the grid when we present our results to reduce the impact of numerical edge effects. Our choice of grid size and ranges also provides us with a slightly better resolution of $\Delta\pmb{v} \sim [3, 2, 2]$  km s$^{-1}$.

Since the algorithm itself does not take into account the measured uncertainties of the parameters, we statistically resample every source that we use. To do this we draw alternative parameters for each source from a multivariate Gaussian in $\mu_{\alpha\ast}$, $\mu_\delta$, and $\varpi$, with the measured values as means and the uncertainty and correlation coefficients in the covariance matrix. We ignore the uncertainties on RA and Dec as they are negligible. By comparing the inferred velocity distribution of these resamples with the original sample we can estimate how significant the observed features are. We  find no significant deviations from the initial sample. One example using a resampled distribution can be seen in Appendix \ref{app:resamples}.

\section{Sample selection}\label{sec:sample}

\begin{figure}
	\centering
	\includegraphics[width=0.40\textwidth]{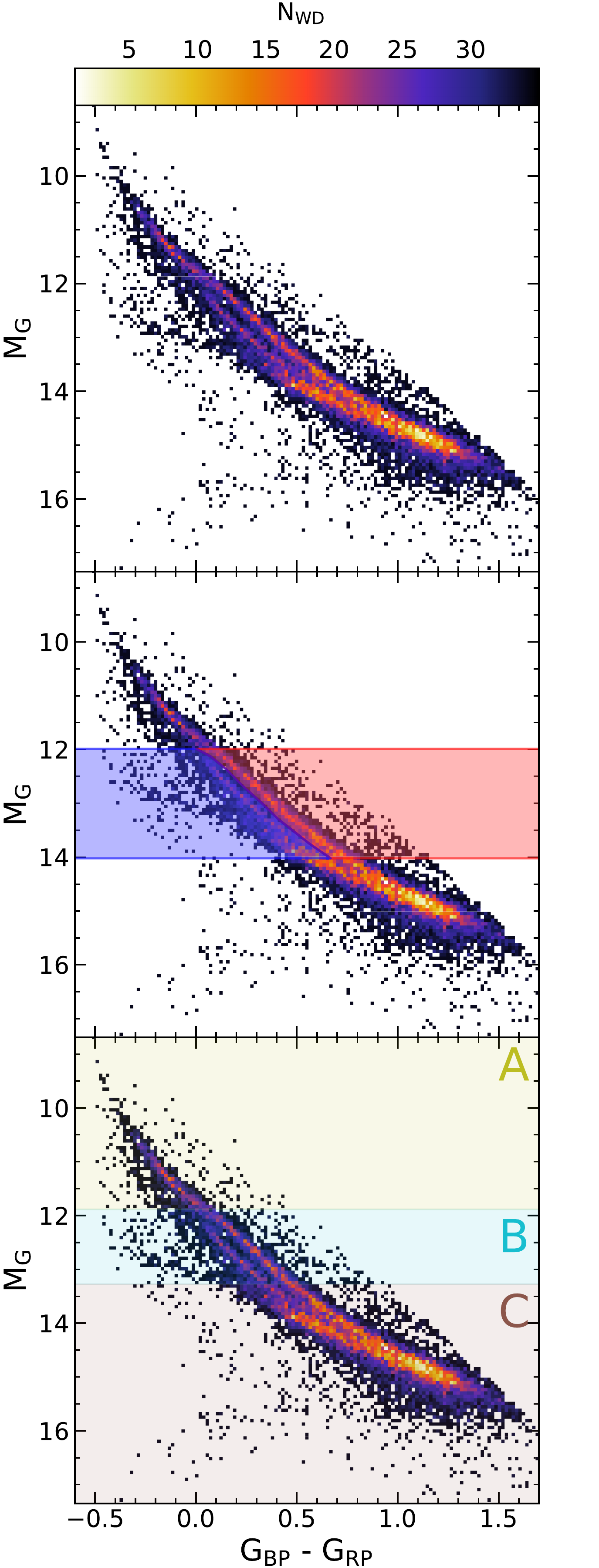}
	\caption{\textit{Top}: Colour-Magnitude diagram for the white dwarfs within 100 pc. Colour shows density of WDs on a 150x150 grid. \textit{Middle}: The red and blue selections used to split across the bifurcation. The vertices of the two regions can be seen in appendix \ref{app:regions}. \textit{Bottom}: Three different bins in absolute magnitude shown on top of the CMD.}
	\label{fig:wd_cuts}
	\vspace{-10pt}
\end{figure}
\begin{table}
	\centering
	\caption{The names of the various samples used and the number of sources in them}
	\begin{tabular}{l | r | l}
		Name & $N_\mathrm{WD}$ & Description\\
		\hline
		{\tt all\_500} & 129 675 & WDs with $d < 500$ pc  \\
		{\tt all\_200} & 54 330 & WDs with $d < 200$ pc \\
		{\tt all\_100} & 14 985 & WDs with $d < 100$ pc\\
		{\tt red\_500} & 32 640 & Red seq. WDs with $d < 500$ pc \\
		{\tt red\_200} & 19 883 & Red seq.WDs with $d < 200$ pc \\
		{\tt red\_100} & 2 909 & Red seq. WDs with $d < 100$ pc \\
		{\tt blue\_500} & 27 809 & Blue seq. WDs with $d < 500$ pc \\
		{\tt blue\_200} & 18 653 & Blue seq. WDs with $d < 200$ pc \\
		{\tt blue\_100} & 2 842 & Blue seq. WDs with $d < 100$ pc \\
		{\tt A} & 43 225 & WDs with 7.4 mag $< M_G <$ 11.9 mag\\
		{\tt B} & 43 225 & WDs with 11.9 mag $< M_G <$ 13.3 mag \\
		{\tt C} & 43 225 & WDs with 13.3 mag $< M_G <$ 18.3 mag \\
	\end{tabular}
\label{tab:samples}
\end{table}
We use the data from Gaia EDR3 \citep{edr3} and select a Solar Neighbourhood sample as this allows us to approximate the velocity distribution as constant across the sample, which is required for the analysis. To achieve this we set a minimum parallax of 2 mas. We apply several quality filters to select a good sample. The re-normalised unit weight error (RUWE) described in \cite{ruwe} is a goodness-of-fit statistic, we require that it is less than 1.15 based on an inspection of the distribution of values in an unfiltered sample and of the CMD. The corrected BP and RP flux excess, $C^*$, is calculated following the procedure laid out in \cite{riello} as part of our ADQL query into the column {\tt excess\_flux} (The full query can be seen in appendix \ref{app:query}). We use their selection criterion and require that
\begin{equation}
	\lvert\mathrm{\tt excess\_flux}\rvert < 3\lvert c_0 + c_1 G^m \rvert,
\end{equation}
where $c_0 = 0.0059898$, $c_1 = 8.817481\times10^{-12}$, and $m = 7.618399$. 

We create the WD sample by setting $M_g > 9.6 + 3.7(G_\mathrm{BP} - G_\mathrm{RP})$. As an additional test, we compare our WD sample to that of \cite{gentile-fusillo} and find that 99.5\% of our WDs are available within their catalog. In their paper they calculate the probability that each source if a WD, $P_{WD}$, and we find that of the cross-matched sample 98\% of stars has $P_{WD} > 0.9$, indicating a high degree of confidence that our sources are WDs.

The WDs are split along the bimodal sequences along a line selected by eye into an upper and lower sequence (hereafter referred to as the red and blue sequences). We limit these cuts to $12 \lesssim M_G \lesssim 14$ where the bifurcation is clearest. All samples and the number of sources in them are listed in Table \ref{tab:samples}. To ensure that we are robust to the specific choice of line, we shift the line $\pm$0.05 mag in colour, creating two different versions of the red and blue samples. We find that our results do not change when we shift the line. For WDs with $d < 100$ pc we show the red and blue sequences on top of the CMD in Fig. \ref{fig:wd_cuts}.

Our selection includes WDs at distances up to $\sim$500 pc. In \textit{Gaia} the nominal brightness limit is $G = 20.7$ \citep{edr3} which in an ideal case would allow \textit{Gaia} to detect sources as faint as $M_G=15.7$ within 100 pc and $M_G=14.2$ within 200pc. In addition to this we place a parallax uncertainty criterion of $\varpi/\sigma_\varpi > 10$ which means that at 100 pc and 200 pc the uncertainties must be smaller than 1 mas and 0.5 mas respectively. In \textit{Gaia} the typical parallax uncertainty at $G=20.7$ for five-parameter solutions is 1.3 mas \citep{edr3}. The brightness and uncertainty limits means that beyond 100 pc we will start to be affected by incompleteness, and therefore Malmquist bias. Conversely our samples within 100 pc is free of this bias, especially the red and blue samples which only go as faint as $M_G=14$.  We also use two WD samples limited in distance to 200 pc and 500 pc which are limited by this bias. The effect of the bias will not be exactly the same for the two sequences as the blue sequence is slightly fainter than the red. However, the shift in magnitude between the sequences is sufficiently small that we can neglect it and assume they share the same bias. This means that any difference we see between the two sequences \textbf{is} not be due to the Malmquist bias as it affects the sequences in effectively the same way. A two-sample KS-test on the velocity dispersion of the samples between 0-100pc and between 100-200pc shows that these are drawn from the same population, so we can be confident that Malmquist bias has not had a significant effect on our sample out to 200 pc (see below).

We also split the sample into three bins in absolute magnitude.  These bins are chosen such that they have an equal numbers of WDs from the {\tt all\_500} sample. The bins are illustrated in Fig. \ref{fig:wd_cuts}, labelled {\tt A}, {\tt B}, and {\tt C} and represent a brighter, intermediate, and faint selection of WDs respectively. As the WDs grow older they will cool and become fainter, so by looking at different magnitudes we should be able to detect age-dependent variation in the velocity distribution.

\section{Velocity dispersion}\label{sec:moments_results}
\begin{figure*}
	\centering
	\includegraphics[width=1\textwidth]{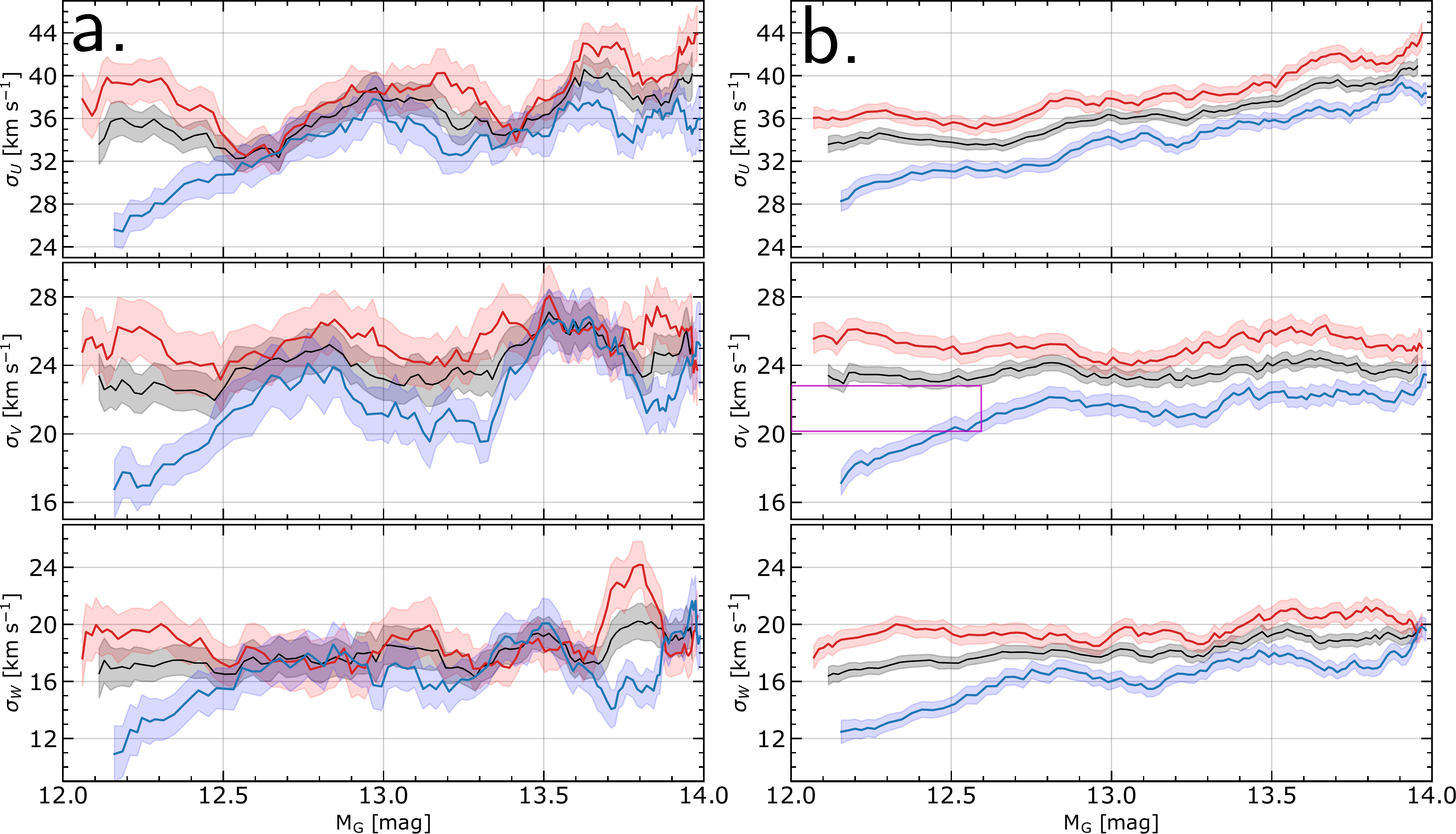}
	\caption{\textit{Panel a.} Dispersions in $U$, $V$, and $W$ calculated for samples {\tt all\_100}, {\tt red\_100}, and {\tt blue\_100} using a moving window in absolute magnitude and shown with black, red, and blue colours respectively. The shaded region shows the 1$\sigma$ uncertainty. The red and blue lines appear to be separate for brighter white dwarfs in all three directions but become mixed towards the fainter end of the sequence. \textit{Panel b.} Same as \textit{a.} but for samples {\tt all\_200}, {\tt red\_200}, and {\tt blue\_200}. For these WDs the split between the red and the blue sequences is much more pronounced and now clearly so at all absolute magnitudes. The red sequence appears to have a larger velocity dispersion in all directions and at almost all absolute magnitudes. }
	\label{fig:mov_disp}
\end{figure*}
\begin{figure}
	\centering
	\includegraphics[width=0.48\textwidth]{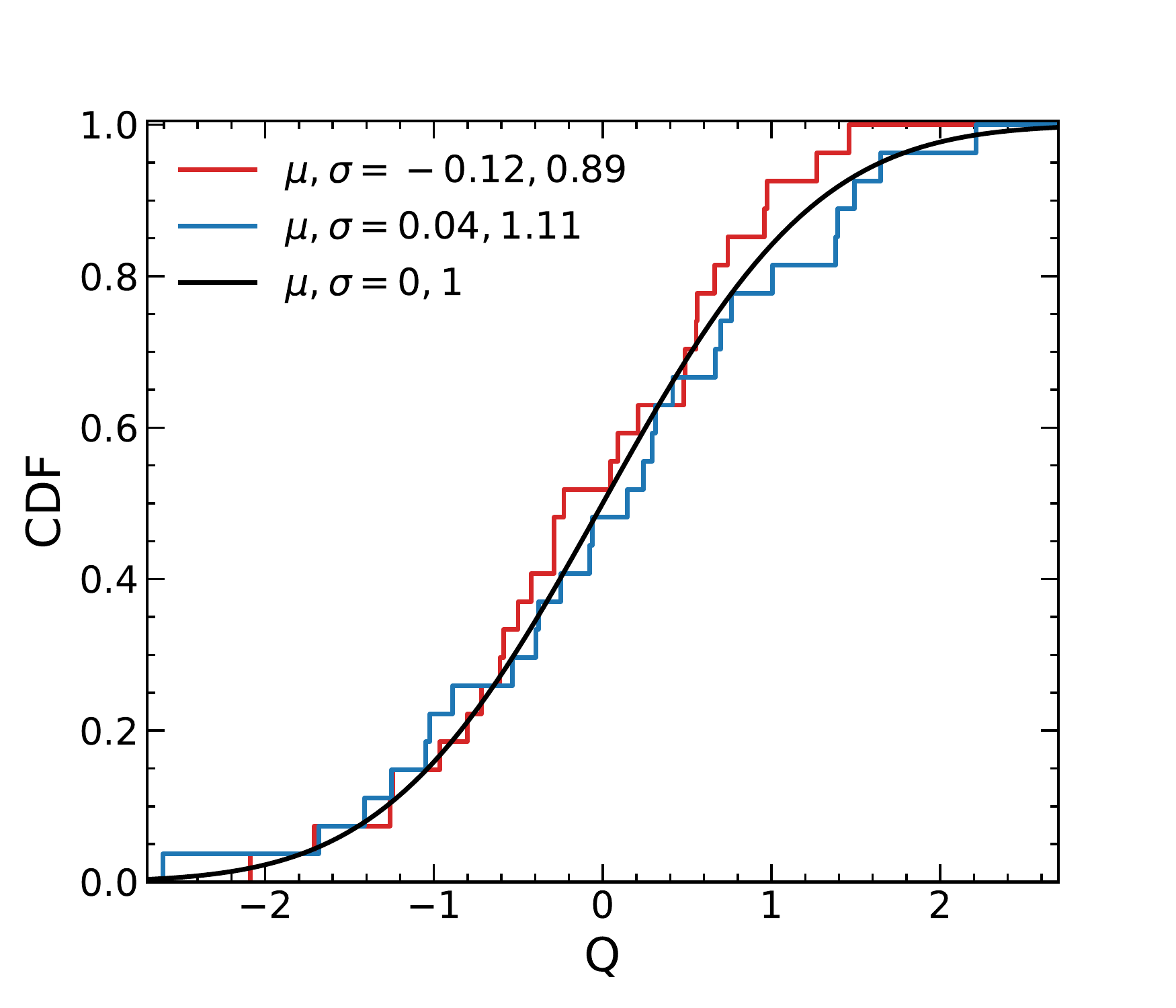}
		\vspace{-10pt}
	\caption{Cumulative distribution of the $Q$-statistic described in the text for the red and blue sequences when comparing samples limited to within 100 pc or between 100-200 pc. The black line shows the Gaussian CDF with $\mu=0$ and $\sigma=1$.}
	\vspace{-10pt}
	\label{fig:cdf}
\end{figure}
\begin{figure*}
	\vspace{-8pt}
	\centering
	\includegraphics[width=1\textwidth]{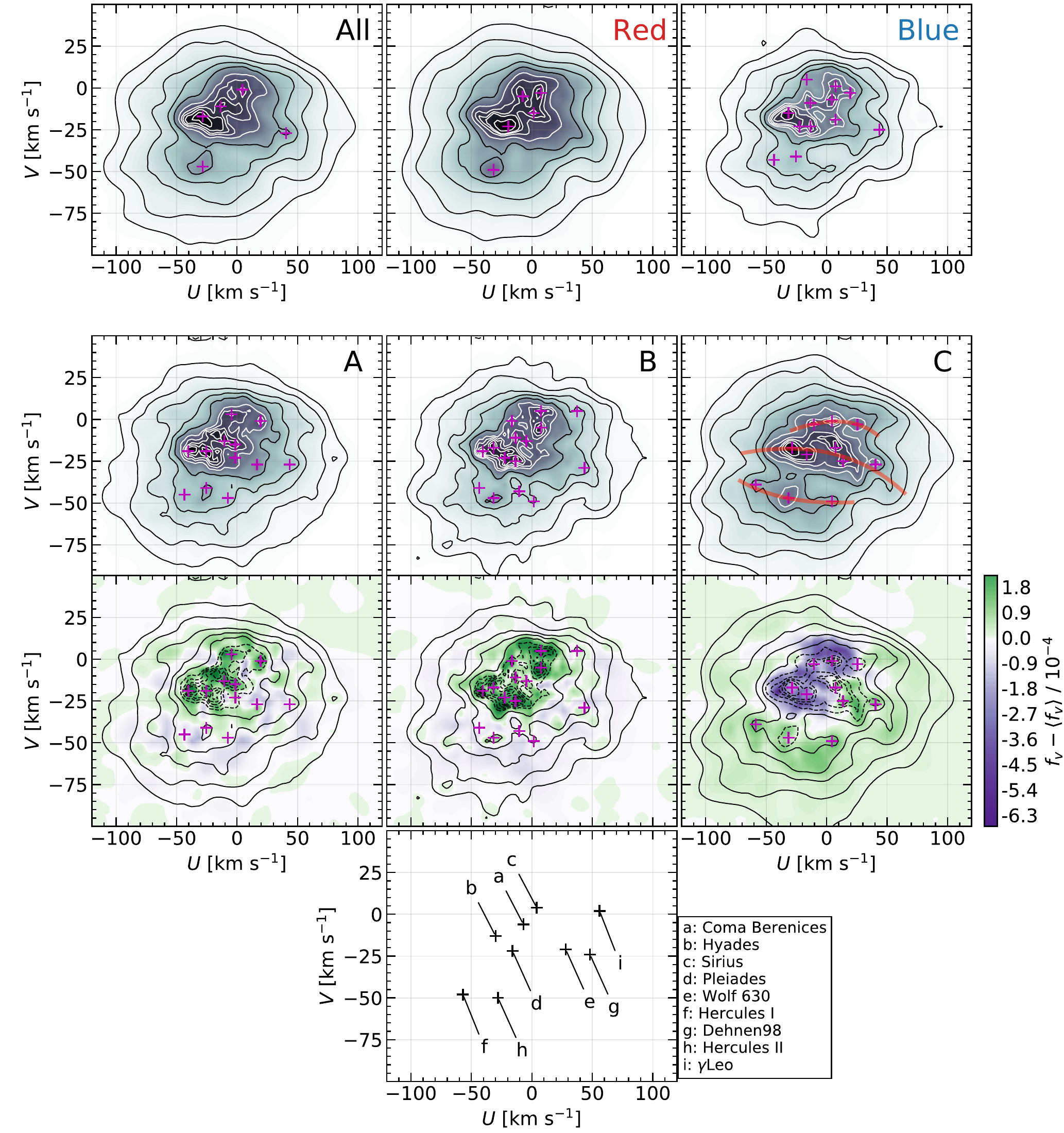}
	\vspace{-10pt}
	\caption{The velocity distribution of WDs in $U$ and $V$. The \textit{top row} shows the distribution for the whole WD sequence and the red and blue sequences corresponding to samples {\tt all\_500}, {\tt red\_500}, and {\tt blue\_500} in Table \ref{tab:samples}. Contour lines are constructed so as to contain 95, 90, 80, 68, 50, 33, 21, 12, 6, and 2 percent of all sources in the sample and magenta crosses show identified peaks with a peak-finding algorithm described in Section \ref{sec:fv_results}. The \textit{second row} contains samples {\tt A}, {\tt B}, and {\tt C} as indicated which are magnitude bins in the full sample. Three horizontal arches are illustrated in sample {\tt C} and explained in section \ref{sec:fv_results}. The \textit{third row} takes the distributions for the three samples {\tt A}, {\tt B}, and {\tt C} and subtracts the mean of all three samples. The contour lines are the same as the individual distributions. The \textit{fourth row} shows the first 9 groups identified in \citet{antoja2012} as black crosses for comparison.}
	\label{fig:UV}
	\vspace{-10pt}
\end{figure*}
\begin{figure*}
	\centering
	\includegraphics[width=1\textwidth]{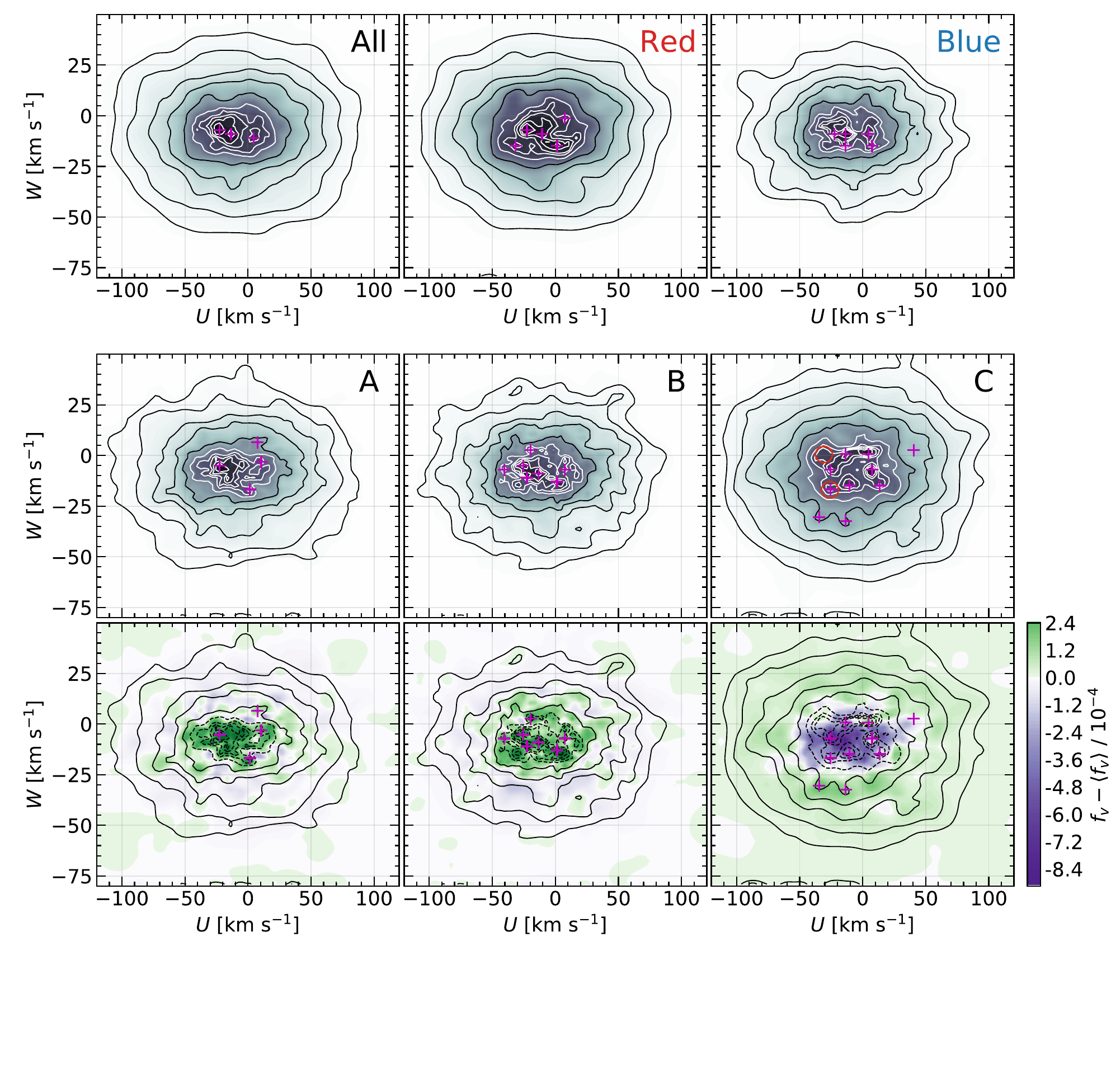}
	\caption{Same as Fig. \ref{fig:UV} but in $U$ and $W$. The circles in panel C marks the peaks that are suggested to be part of a double-peaked feature discussed in section \ref{sec:fv_results}}
	\label{fig:UW}
\end{figure*}
\begin{figure*}
	\centering
	\includegraphics[width=1\textwidth]{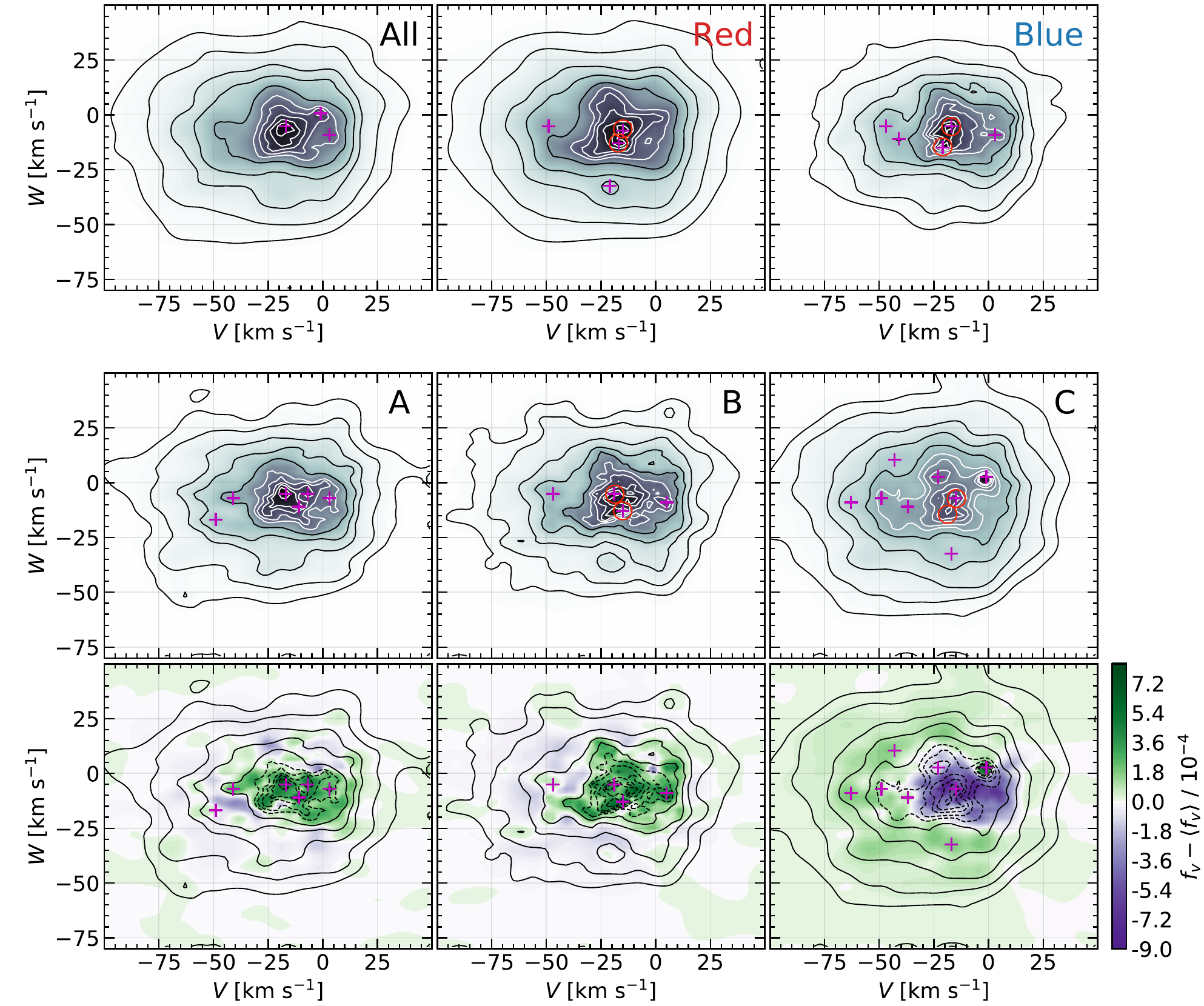}
	\caption{Same as Fig. \ref{fig:UW} but in $V$ and $W$.}
	\label{fig:VW}
\end{figure*}
In section \ref{subsec:moments_theory} we showed how we can calculate the velocity dispersion of a sample of stars using only their positions and tangential velocities. We calculate 3D velocity dispersions for a moving window in absolute magnitude separately for the WD samples limited to 100 pc and 200 pc. We estimate the uncertainty within the windowed sample by calculating the moments for 500 bootstrapped samples. The standard deviation of all of the resulting statistics is used to provide a 1$\sigma$ uncertainty. The complete 100 pc sample's velocity dispersions are seen in panel \textit{a.} of Fig. \ref{fig:mov_disp}. Generally we would expect the velocity dispersion to increase with fainter magnitude due to dynamical heating over time. This is true overall, though the component $\sigma_V$ is approximately constant for the red and black sequences over the range of magnitudes considered.

The red sequence has a larger velocity dispersion than the blue at low magnitudes in all directions to the extent that their 1$\sigma$ regions do not overlap. Beyond this the separation is not as great and harder to make out. In both $U$ and $V$ the median places the red sequence above the blue. For a comparison we look also at the velocity dispersion for all the WDs within 200 pc in panel \textit{b.} in which the separation is very clear at all magnitudes and for all three velocity components. The 1$\sigma$ regions of the red and blue barely overlap with the joint sample. To determine whether or not the two distance-limited samples probe the same underlying distribution, we calculate the statistic
\begin{equation}
	Q = \frac{\sigma_{A} - \sigma_{B}}{\sqrt{\Delta\sigma_{A}^2 + \Delta\sigma_{B}^2}},
\end{equation}
where $\sigma_{A,B}$ is the velocity dispersion for samples A or B and $\Delta\sigma_{A,B}$ their uncertainties estimated through bootstrapping. We calculate Q for both sequences with A being the 100 pc limited samples and B selected to be the stars between 100-200 pc to avoid overlapping sources. We calculate Q for a number of binned points in $M_G$ and if the two distance-limited samples are drawn from similar underlying distributions the distribution of $Q$ should be Gaussian with $\mu=0$ and $\sigma=1$. The cumulative distribution of $Q$ can be seen in Fig. \ref{fig:cdf} for the red and blue sequences and shows good alignment with the analytical Gaussian CDF. In addition to this we perform a Kolmogorov-Smirnov (KS) test between the different samples and the Gaussian CDF under the null hypothesis that they arise from the same distribution. The KS testing yields a $p$-value of 0.936, 0.834, and 0.936 when testing the red/blue, red/Gaussian, and blue/Gaussian Q distributions respectively, clearly demonstrating that they arise from the same Gaussian distribution and therefore the difference between our two samples is simply due to statistical noise. In summary, there is a statistically significant separation in the velocity dispersion between the red and blue sequences.

If the bifurcation is caused by a bimodal WD mass distribution, this could be explained by the progenitors of the WDs in the two different sequences. The lower mass WDs will come from less massive main-sequence stars, which would mean that these progenitors will have been dynamically heated for a longer duration of time and the lower mass WDs would be born with larger dispersions.

The separation between red and blue sequences is larger at brighter magnitudes, which corresponds to younger WDs. This can be expected if the blue sequence comes from massive WDs which are younger than their red sequence, less massive, counterparts. Most of the dynamical heating occurs during the first few Gyr of a stars life (e. g., \citealt{nordstrom, galacticdynamics}) and so would already have occurred for the red sequence WDs but still be occurring for the blue sequence ones. Over time as WDs on both sequences cool to fainter magnitudes their heating tracks would align to become more parallel, which is visible in our figures.

Undeniably the bifurcation in the CMD is well-described by a difference in the atmospheric composition of WDs \citep{bedard}. It is however difficult to reconcile with the observed difference in the kinematics that we display here which requires significantly different dynamical heating histories between the two sequences. If there is a significant age difference between the two populations however this would result from the expected heating from the Galactic disk. We can estimate the age difference of stars in the two sequences using the age-velocity dispersion relationship from \cite{aumerbinney} and applying it to the total velocity dispersion of the two samples, \texttt{red\_500} and  \texttt{blue\_500}. This suggests a typical age of  {$\sim$}7.7 Gyr and {$\sim$}4.9 Gyr respectively for the red and blue sequences. We are unaware of any mechanism that could explain these kinematics purely by a difference in atmospheric composition.

As mentioned in Section \ref{sec:sample} we also selected WDs out to $\varpi = 2$ mas or 500 pc. Beyond 200 pc the separation between red and blue sequences remains the same and did not provide any further insights.


\section{Velocity distribution of WDs}\label{sec:fv_results}

We apply the maximum penalized likelihood estimate to all of the samples to produce a full 3D velocity distribution for which we show the projections in the planes $U$-$V$, $U$-$W$, and $V$-$W$. To identify where the peaks of the distribution are we apply the \textit{peak\_local\_max} function available as part of the \textit{scikit-image}\footnote{\url{https://scikit-image.org/}} Python package \citep{scikit-image}. The results in the $U$-$V$ plane are presented first in Fig. \ref{fig:UV}. In the first row we compare the velocity distribution of all our WDs with those of WDs identified as belonging to the red and blue sequences as laid out in Section \ref{sec:sample}. It is immediately clear that the WD velocity distribution is overall very similar to the rest of the Solar Neighbourhood (e.g., \citealt{katz}) which is not too surprising considering that the WDs are older stars of the same population.

We can identify a few familiar moving groups such as the \textit{Hyades}, \textit{Pleiades}, and \textit{Hercules} across all three samples very clearly. To a lesser extent we can also identify higher velocity moving groups beyond $U > 25$ km s$^{-1}$ such as \textit{Wolf 630} and \textit{Dehnen98}.
\begin{table*}
	\centering
	\caption{Identified features in the velocity distributions at approximate coordinates. We omit the velocity in $W$ when it cannot be determined accurately. Filled or hollow circles are for clear and weak features respectively. Group 5 has an asterisk indicating unclear membership in the moving group which is further discussed in Section \ref{subsec:fv_discussion}.}
	\begin{tabular}{lrrrccccccl}
		\hline
		Feature & $U$ & $V$ & $W$  & {\tt all\_500} & {\tt red\_500} & {\tt blue\_500} & {\tt A} & {\tt B} & {\tt C} & Moving Group   \\ \hline
		1\dotfill  & 4     & -1         & \ldots & \newmoon  & \newmoon & \newmoon  &                       & \fullmoon     & \newmoon  & Sirius         \\
		2\dotfill  & -13   & -11    & -10      & \newmoon  & \fullmoon    & \newmoon    & \newmoon  & \newmoon  &                    & Coma Berenices \\
		3\dotfill  & -28   & -17   & \ldots  & \newmoon  &  \fullmoon   & \newmoon  & \newmoon  & \newmoon  & \newmoon & Hyades         \\
		4\dotfill  & -28   & -47  & \ldots  & \newmoon  & \newmoon & \newmoon  & \newmoon  & \newmoon  & \newmoon & Hercules    \\
		5\dotfill  & 7    & -19      & \ldots  & \fullmoon &                   &  \newmoon   &                      &                     & \newmoon& Coma Berenices$^\ast$               \\
		6\dotfill  & 1     & -48     & \ldots  & \fullmoon    &                   & \fullmoon    & \fullmoon    & \fullmoon   & \fullmoon   & Hercules                \\
		7\dotfill  & 1    & -15     & \ldots  &                     & \newmoon  &                     & \newmoon   &                   &  \fullmoon  & Coma Berenices               \\
		8\dotfill  & 16    & -26    & \ldots  &                    &                   &                     & \fullmoon    &                    & \newmoon & Wolf 630       \\
		9\dotfill  & -19   & -23   & -8        & \fullmoon  & \newmoon & \newmoon   & \newmoon  & \newmoon  & \newmoon  & Pleiades       \\
		10\dotfill  & 40    & -27  & \ldots  & \newmoon &  \fullmoon    & \newmoon & \newmoon  &\newmoon   & \newmoon & Dehnen 98      \\
		11\dotfill & 19    & -1       & \ldots  &                   &                    &  \newmoon  & \newmoon  &                    & \fullmoon   & Sirius               \\
		12\dotfill & -7    & -5       & \ldots  &                   &  \newmoon  &                    &                    &                    & \fullmoon   & Coma Berenices
	\end{tabular}
	\label{tab:features}
\end{table*}
The moving groups in the blue sequence are more narrowly distributed around their group mean overall than those of the red sequence, where the groups are spread out more across the central regions in the space due to the overall larger velocity dispersion shown in Section \ref{sec:moments_results}.

We show the three samples of different absolute magnitudes in the second row of Fig. \ref{fig:UV}, going from brightest ({\tt A}) to faintest ({\tt B}). For the individual distributions we see the same type of structure as we did for the whole sample. It appears as though there is more structure in these subsamples than the former three. However, as the samples contain less stars than the {\tt all\_100} sample but use the same smoothing parameter $\alpha$ it has a noisier distribution. The fainter sample, {\tt C}, has older stars and thus a larger velocity dispersion as expected which `smears' out the distribution which reveals an arch-like structure in {\tt C}, with three horizontal arches at $V$ of 0 km s$^{-1}$, -20 km s$^{-1}$, and at -40 km s$^{-1}$ which are illustrated in the plot. This type of structure has been shown to exist in the Solar Neighbourhood \citep{katz} and can be attributed to dynamical resonances with the spiral arms and Galactic bar \citep{trick1}.

The third row takes the three samples {\tt A}, {\tt B}, and {\tt C}, and subtracts the mean of all three to emphasise the parts of the distribution that are shared between the samples or are independently significant. Our expectation is that the bright, young sample with low velocity dispersion ({\tt A}) should be over-abundant near the centre of the distribution and the faint sample should show the opposite. Stars with very large $|V|$ belong to populations with their guiding centres at significantly larger/smaller Galactocentric radii and would be visiting the Solar Neighbourhood when their orbits are sufficiently dynamically heated. Thus, they would be older and part of a fainter sample. This effect can be seen in the third row where samples {\tt A} and {\tt B} dominate the centre of the distribution whilst {\tt C} is represented more in the wings with, for example, the \textit{Hercules} moving groups being overabundant. Asymmetric drift also causes the mode of the distribution to shift towards more negative $V$.

The region around $(U, V) \approx (7, -19)$ km s$^{-1}$ is underdense in {\tt A} and {\tt B} but surprisingly is over-abundant in {\tt C}, with an identified peak in that region as well. A review of recent articles that investigate the velocity distribution in $U$-$V$ for the Solar Neighbourhood reveals that \cite{antoja2012} and \cite{katz} do not identify a known moving group in this region while \cite{kushniruk} does and attributes it to the Coma Berenices stream.

The distributions in $U$-$W$ and $V$-$W$ in Fig. \ref{fig:UW} and Fig. \ref{fig:VW} are not as rich in structure as the $U$-$V$ plane, but we can, however, identify \textit{Coma Berenices} and \textit{Pleiades} by comparing the two planes. Our choice of smoothing parameter $\alpha$ is chosen to fit best with the sample {\tt all\_500} and hence will be under-estimated for the smaller samples ({\tt red\_500} and {\tt blue\_500}) and as such the features seen in them may disappear with an $\alpha$ that is appropriate for smaller samples. A feature identified by \cite{wd98} is a double-peaked feature along $W$ in the $U$-$W$ plane at $U \approx -30$ km s$^{-1}$ and in the $V$-$W$ plane at $V \approx -20$ km s$^{-1}$. This double-peaked feature is very vaguely present in $U$-$W$ for sample {\tt C}. In the $V$-$W$ plane however we can identify the feature clearly in both {\tt red\_500} as well as {\tt blue\_500}, and perhaps vaguely in {\tt B} and {\tt C}. We mark the proposed double-peak feature with circles around the involved features. This seems to imply that the double-peaked feature, of which the \textit{Pleiades} appears a part of, is limited to fainter, possibly older stars.

In Table \ref{tab:features} we summarise the features we can see and mark whether we can see them clearly, somewhat, or not at all.  We also list their locations in all three velocity dimensions (see for comparison Table 2 from \citealt{wd98}). For most of the moving groups we are able to associate a previously known one following \cite{kushniruk} but no known group accurately describes feature 5 in the table, meaning that it may be a new feature. Compared to previous results the features presented here are located mostly at lower velocities ($|v| < 50$ km s$^{-1}$). It may be that the WDs in groups at larger velocities are not numerous enough to appear in the distributions or that their dispersion is too great.

\subsection{The stellar warp}\label{sec:warp}
\begin{figure}
	\centering
	\includegraphics[width=0.5\textwidth]{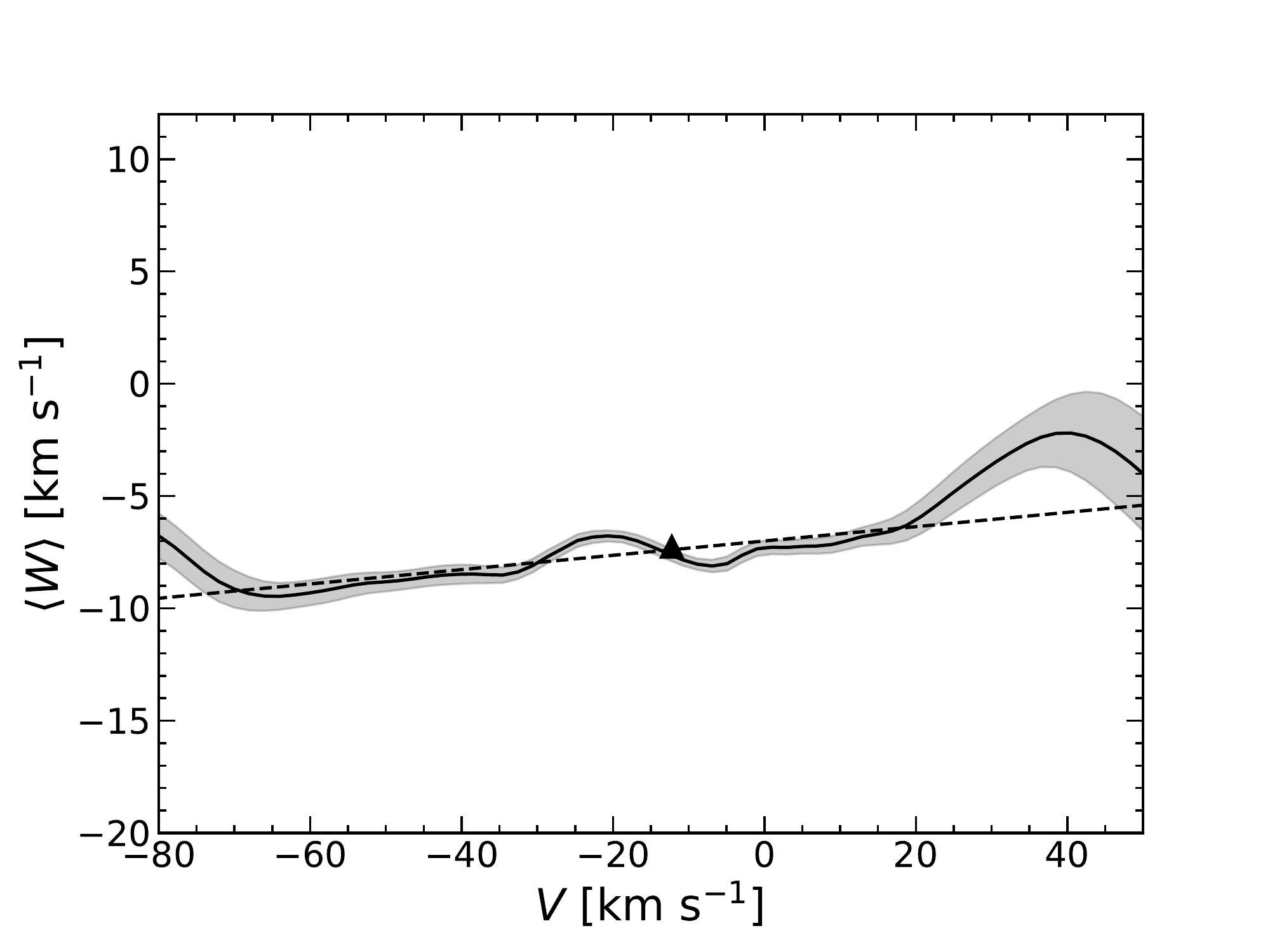}
	\caption{Mean of the vertical motion, $\langle W\rangle$, in the central regions of $f(V, W)$ for sample {\tt all\_500}. The location of the LSR is marked by a triangle and is taken from \citet{LSR}. The errorbars show the $1\sigma$ Poisson noise. The dashed black line shows a weighted linear fit.}
	\label{fig:warp}
\end{figure}
In his analysis of \textit{Hipparcos} data, \citet{wd98} found evidence for the stellar warp by investigating $\langle W\rangle$ as a function of $V$. For $V > 10$ km s$^{-1}$ the velocity distribution was skewed towards positive $\langle W\rangle$. We perform the same analysis here of our largest WD sample {\tt all\_500} and the results can be seen in Fig. \ref{fig:warp}.  To do this we require a larger grid of $\pmb{n} = [200, 152, 152]$ covering $U \in [-200, 200]$ km s$^{-1}$, $V \in [-150, 150]$ km s$^{-1}$, and $W \in [-150, 150]$ km s$^{-1}$ to ensure that none of the edge effects discussed in Section \ref{subsubsec:maximizing} affect the results. We find that our results also show increasing $\langle W\rangle$ with increasing $V$.

We compare our results to \cite{sd18} who use \textit{Gaia}-TGAS to determine the average vertical velocity as a function of azimuthal velocity and angular momentum for stars in the two cones towards the Galactic centre and anti-centre with angular radius of 30$^\circ$ (their Fig. 6). Since our samples are in the Solar Neighbourhood the velocity in $V$ is a proxy for angular momentum since $L_z = -R(V_\odot + V)$. Just as in our sample, the average vertical velocity increases with angular momentum and shows a dip just beyond the Local Standard of Rest (LSR). At large angular momentum, corresponding to $V > 15$ km s$^{-1}$ in the Solar Neighbourhood there a significant increase in $\langle W\rangle$ for all of their samples as well as ours. Towards  $V = 50$ km s$^{-1}$ $\langle W\rangle$ decreases again which is consistent with a similar decrease at higher angular momentum seen by \cite{sd18}.

More recently, \cite{anticentre} also looked at the vertical velocity profile of stars outside the solar radius as a function of angular momentum. We see this in their fig. 11 which shows this profile for several different stellar types, of which the young main-sequence stars are the ones to show a decrease in vertical velocity at $L_z^\ast \sim 2400$ kpc km s$^{-1}$ corresponding to around $V = 60$ km s$^{-1}$. However, our sample and the \textit{Gaia}-TGAS sample are in the Solar Neighbourhood where stars with such large angular momentum must have large radial excursions to be included in the sample, and therefore are very likely to be old. This is in contrast to the stars in \cite{anticentre} which are young and located beyond 10 kpc from the Galactic centre. While the decrease in the vertical velocity profile we see could therefore be related to these decreases seen in other datasets, we note that the decrease is still within 1$\sigma$ uncertainty, and therefore the most we can say is that the results are mutually consistent.

For $V < 20$ km s$^{-1}$ where the Poisson noise is not as great, the average velocity, $\langle W\rangle$, has a relatively small tilt. We perform a weighted linear fit for $V \in [-80, 50]$ km s$^{-1}$ and find a slope of $\sim$0.03 (shown in Fig. \ref{fig:warp}), which is somewhat larger than $\sim$0.02 found by \cite{sd18}. If we instead restrict our fit to the relatively well constrained region $V \in [-80, 20]$ km s$^{-1}$ we find a slope of $\sim$0.025 in agreement with the previous results.

\section{Discussion}\label{sec:disc}

\subsection{The bifurcated WD CMD}
In Section \ref{sec:intro} we reviewed the literature on the nature of the bifurcation of the CMD in the \textit{Gaia} data. We know that there exists a bimodality in the mass distribution of spectral class DA WDs \citep{kilic18, jimenez19, kilic2020} at around {$\sim$}0.6 M$_\odot$ and {$\sim$}0.8 M$_\odot$.

Cooling tracks for these masses of DA WDs would produce a bifurcation that appears very similar to the one visible in \textit{Gaia} \citep{el-badry18}. However, massive WDs will form sooner and begin cooling at an earlier time than the less massive ones, which means that for these cooling tracks to be simultaneously visible in the CMD the massive WDs must be a) formed through the merger of lower mass WD binaries, b) formed at a later time than the less massive WDs, or c) cooling slower due to crystallization. Scenario a) was suggested by \cite{kilic18} but subsequently dismissed by \cite{kilic2020} due to merger models being unable to produce a mass distribution that fit the observations as well as, perhaps more importantly, not being able to find a significant amount of young and massive DA WDs with high velocities. This leaves scenarios b) and c). In the first of these scenarios a multimodal age distribution could produce massive WDs at a later time for a bimodal mass distribution which would have a smaller velocity dispersion as they have not had as much time to be heated. In the latter scenario, crystallization \citep{tremblay19} will result in a slowdown of the cooling of massive WDs which means the massive sequence can consist of a mixture of young and older massive WDs. Both of these scenarios would cause the massive WDs to have a lower velocity dispersion. In our Fig. \ref{fig:mov_disp} it is clear that this behaviour is observed. The velocity distributions for the red and blue sequences in Figures \ref{fig:UV}, \ref{fig:UW}, and \ref{fig:VW} exhibit this as well with the red sequence showing a larger velocity dispersion.

Atmospheric composition can elegantly explain the visible bifurcation in the \textit{Gaia} data with an upper DA branch and a lower branch of He-dominated WDs with trace amounts of H or other metals. The bimodal WD mass distribution fits well with this explanation when described by the process of crystallisation or by the inclusion of young massive WDs. However, atmospheric composition alone does not provide the difference in kinematics between the two sequences that we identify. This difference arises naturally however, with a bimodal WD mass distribution from a multimodal age distribution. 

The cooling tracks of {$\sim$}0.6 M$_\odot$ and {$\sim$}0.8 M$_\odot$ DA WDs would also lie in the region where the bifurcation exists (e.g, \citealt{bergeron}). We cross-match our \texttt{red\_100} and \texttt{blue\_100} with the MWDD to determine what fractions of stars are DA or non-DA and find that around 85\% of the red sample cross-match and 39\% of the blue sample cross-match are DAs. The red cross-match shows that the this sequence is likely to be comprised of older, less massive DAs. The DAs in the blue-cross match have higher mass with a mean {$\sim$}0.8 M$_\odot$ as expected in keeping with previous results (e. g, \citealt{kilic2020}).
	
The bifurcation seen in \textit{Gaia} clearly has contributions from both atmospheric differences as well as different WD mass cooling tracks. It remains unclear whether the origins of the massive WDs is recent bursts of star formation or a pile-up of WDs with a mixture of ages due to delayed cooling from crystallisation, or a mixture of both. In \cite{fantin} the star formation history of the Galactic disk is investigated using WDs and suggests star formation increases at $3.3\pm 1.8$ Gyr ago and is roughly constant for ${\sim}5$ Gyr prior to that. A detailed study of the age distribution of the WD population using accurately determined ages, masses, and spectral types would provide valuable insight into these questions and analytical modelling of WD formation and evolution following bursts of star formation  would be a good avenue to test the formation avenues of these WDs.

\subsection{The velocity structure of WDs}\label{subsec:fv_discussion}
 For the first time, we present the velocity distribution of WDs in the Solar neighbourhood in addition to their velocity moments. We find that the WD velocity distribution in $(U, V)$ shares many features with the velocity distribution of main-sequence stars when comparing our results to recent maps of the kinematic structure of the Solar neighbourhood like those of \cite{katz}, \cite{kushniruk}, or \cite{antoja2012}. When we separate our sample along this bifurcation we can see very similar velocity distributions, with the notable differences being the red sample having a larger velocity dispersion.

The CMD is also split into three equally sized samples based on their absolute magnitude. The mean velocity distribution of all three subsamples is subtracted from each individual subsample which reveals an unexpected over-density for the faintest sample, {\tt C}, in the region $(U, V) \approx (7, -19)$  km s$^{-1}$. The location does not match conclusively to any of the known moving groups and is only identified in \cite{kushniruk} who attributes it to be a part of the \textit{Coma Berenices} moving group along with two other identified groups. If this feature is limited to fainter, older stars it could suggest its origin is dynamical.  In \cite{coma_b} it is shown that the \textit{Coma Berenices} moving group is not vertically phase mixed and is localized to negative $b$ only, and it is suggested to be due to a recent passing by a dwarf galaxy such as \textit{Sagittarius} which fits well with passages suggested in literature (for a summary see the lower panel of Fig 2 of \citealt{ruiz-lara}).

The double-peaked feature identified in $W$ is limited to fainter stars which should be part of an older sample. The feature is symmetrical around the average vertical velocity which suggests that this feature might be in the brighter sample as a single feature around the mode containing younger, less dynamically heated stars.

Beyond smaller scale structure we find that the velocity distribution is very similar to that of main sequence stars which can be expected since WDs are subjected to the same dynamical processes as other stars. The distributions also show very clear arches like those seen in dynamical studies of main sequence stars (e.g., \citealt{trick1}). To further narrow down the origin of the observed features requires comparison with model predictions or investigating the ages and abundances of their associated stars. For example, It would be possible to investigate the age evolution of dynamical features in the distributions by studying their appearance for several absolute magnitude bins and associating the absolute magnitude with ages using WD cooling tracks.

\section{Conclusions}\label{sec:conc}

We present the velocity distributions of WDs in the Solar neighbourhood which, to our knowledge, has not been done previously. The velocity distributions are estimated through a penalized maximum likelihood. The data we use comes from \textit{Gaia} EDR3 and is filtered to select a clean and unbiased sample of WDs. We split the WD CMD  across the established bifurcation and into several absolute magnitude bins and find that velocity distribution is similar to that of main sequence stars from previous studies and displays many well-known moving groups. We also identify a novel structure located at $(U, V) \approx (7, -19)$  km s$^{-1}$  which appears only in bins that are fainter or possibly older, and discuss possible explanations for the feature. We also identify a double-peaked feature in $W$, previously established in the \textit{Hipparcos} data by \citet{wd98}, involving mostly fainter stars.

We also explore the mean velocities and velocity dispersions as a function of absolute magnitude and compare it between the two established sequences in the \textit{Gaia} WD CMD. We find that the brighter sequence has larger velocity dispersion than the faint one across all magnitudes: the sequences are two separate kinematic populations. This cannot be explained if the origin of the second sequence is due to WD binary mergers or solely atmospheric composition. Our results are consistent with the observed WD bimodal mass distribution with a multimodal age distribution.

The results of our study sheds light on the bifurcation in the Gaia WD CMD and explores the possibility of accessing the majority of sources in \textit{Gaia} which lack radial velocities. We plan to use EDR3 to investigate the velocity distribution of the Solar neighbourhood using as many stars as we can include to probe the kinematic structure of Milky Way for new insights.





\section*{Data availability}
All data analysed in this paper are publicly available from the Gaia EDR3 archive (\url{http://gea.esac.esa.int/archive/}). The 3D probability distributions used in Figures \ref{fig:UV}, \ref{fig:UW}, and \ref{fig:VW} are available upon request from the corresponding author.
\section*{Acknowledgements}
 This work has made use of data from the European Space Agency (ESA) mission {\it Gaia} (\url{https://www.cosmos.esa.int/gaia}), processed by the {\it Gaia}  Data Processing and Analysis Consortium (DPAC,  \url{https://www.cosmos.esa.int/web/gaia/dpac/consortium}). Funding for the DPAC has been provided by national institutions, in particular the institutions participating in the {\it Gaia} Multilateral Agreement. This research made use of Astropy,\footnote{http://www.astropy.org} a community-developed core Python package for Astronomy \citep{astropy:2013, astropy:2018}. 
 
 We thank members of Lund Observatory for helpful comments and ideas. We give special thanks to Ross Church for constructive discussions and comments. Computations for this study were performed on equipment funded by a grant from the Royal Physiographic Society in Lund. PM is supported by project grants from the Swedish Research Council (Vetenskapr\aa det Reg: 2017-03721, 2021-04153). DH and PM gratefully acknowledge support from the Swedish National Space Agency (SNSA Dnr 74/14 and SNSA Dnr 64/17).
\bibliographystyle{mnras}
\bibliography{references}


\appendix
\newpage
\onecolumn
\section{Gaia archive query}\label{app:query}
The following query has been used on the Gaia archive\footnote{\url{https://gea.esac.esa.int/archive/}} and is detailed in Section \ref{sec:sample}:\\ \\
\begin{lstlisting}[language=SQL, deletekeywords={DEC}]
select bp_rp, phot_g_mean_mag, phot_bp_rp_excess_factor, ra, dec, parallax, pmra, pmdec,
if_then_else(
	bp_rp > -20,
	to_real(case_condition(
		phot_bp_rp_excess_factor - (1.162004 + 0.011464* bp_rp 
																+ 0.049255*power(bp_rp,2) 
																- 0.005879*power(bp_rp,3)),
		bp_rp < 0.5,
		phot_bp_rp_excess_factor - (1.154360 + 0.033772* bp_rp 
																+ 0.032277*power(bp_rp,2)),
		bp_rp >= 4.0,
		phot_bp_rp_excess_factor - (1.057572 + 0.140537*bp_rp)
	)),
	phot_bp_rp_excess_factor
) as excess_flux 
from gaiaedr3.gaia_source
where parallax > 2
and parallax_over_error > 10
and ruwe < 1.15
\end{lstlisting}
\twocolumn

\section{Red and blue WD sequence selection}\label{app:regions}
Table \ref{tab:regions} lists the vertices for the intersect between red and blue regions of the WD CMD outlined in Section \ref{sec:sample}.
\begin{table}
	\label{tab:regions}
	\centering
	\caption{Vertices for the regions in  $M_G$ and $G_\mathrm{BP}-G_\mathrm{RP}$ that make up our red and blue WD sequences.}
	\begin{tabular}{l l}
		$G_\mathrm{BP}-G_\mathrm{RP}$ & $M_G$\\
		\hline
		mag & mag\\
		\hline
		0.668 & 14.021\\
		0.561 & 13.746\\
		0.407 & 13.291\\
		0.323 & 12.982\\
		0.231 & 12.673\\
		0.150 & 12.364\\
		0.071 & 12.124\\
		0.001 & 11.987
	\end{tabular}
\end{table}

\section{Resampled velocity distributions}\label{app:resamples}
At the end of Section \ref{subsec:fv_theory} we explained how we used a statistical resample of our proper motions and parallaxes with their measured uncertainties to estimate the effect that the uncertainties might have. As they are not significant we omit them from the main paper and show a single example of one of these resamples here.
\begin{figure*}
	\centering
	\includegraphics[width=0.9\textwidth]{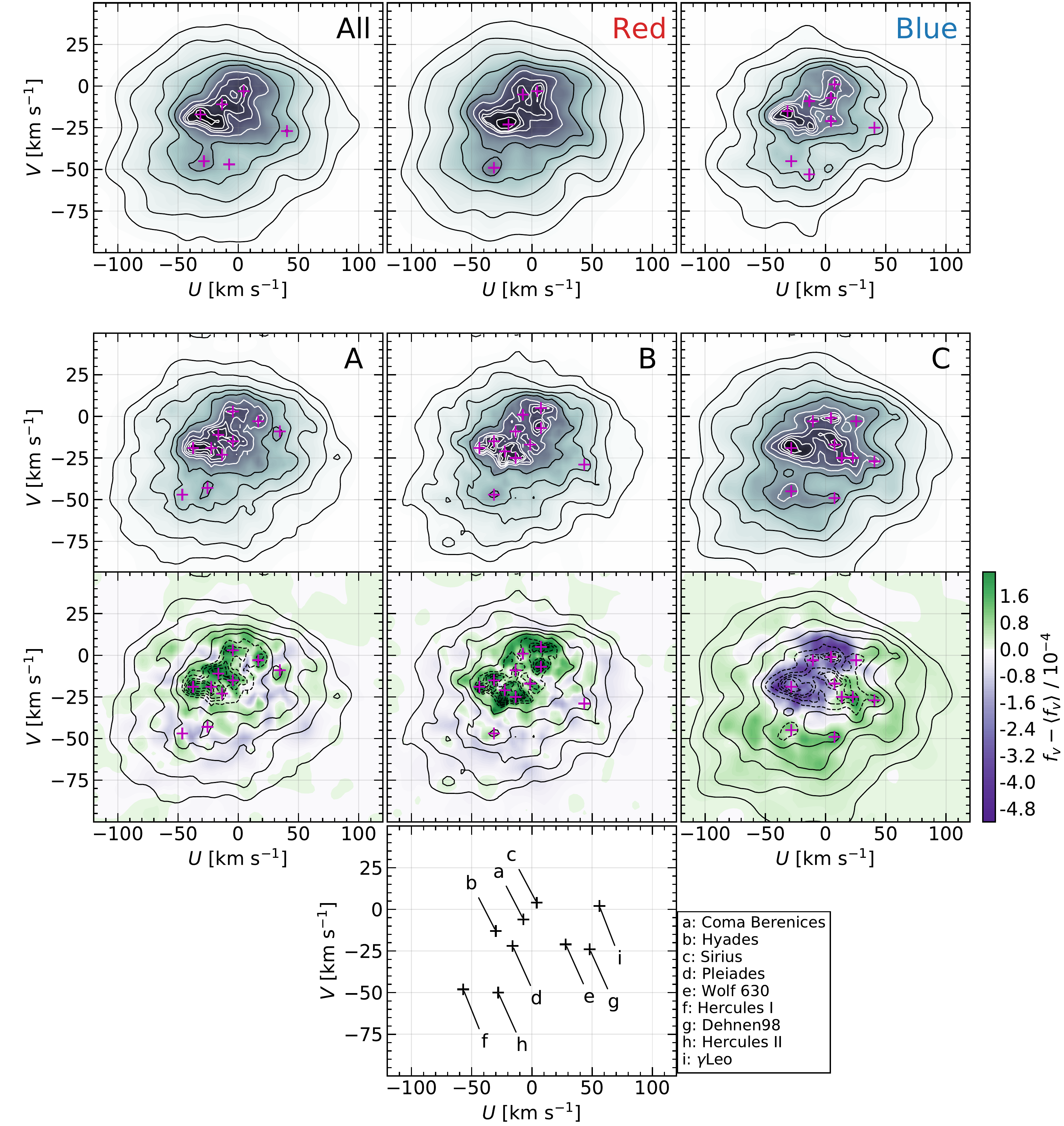}
	\caption{Same as Fig. \ref{fig:UV} using a statistical resample of the proper motions and parallaxes.}
	\label{fig:resample_UV}
\end{figure*}
\begin{figure*}
	\centering
	\includegraphics[width=0.9\textwidth]{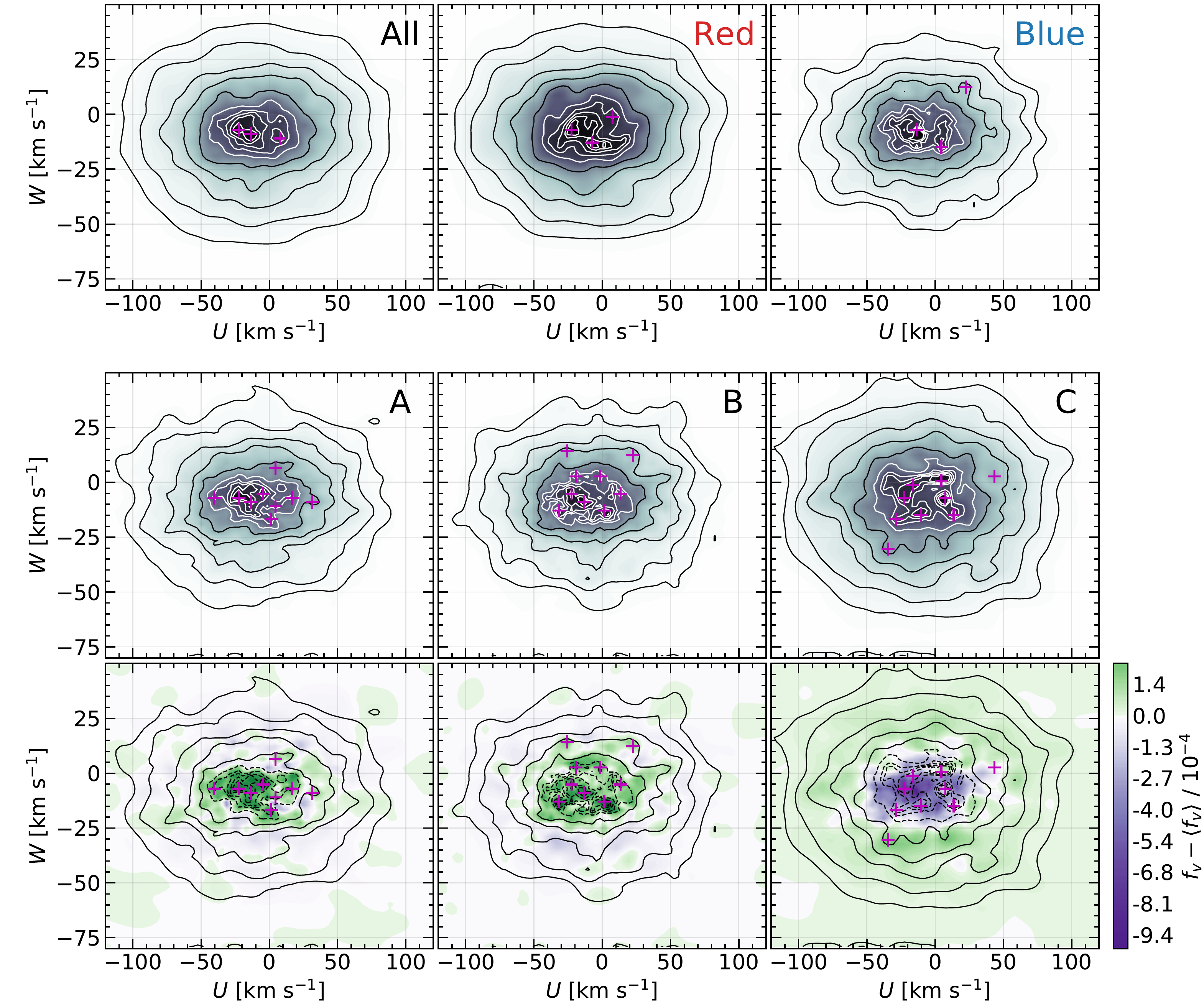}
	\caption{Same as Fig. \ref{fig:UW} using a statistical resample of the proper motions and parallaxes.}
	\label{fig:resample_UW}
\end{figure*}
\begin{figure*}
	\centering
	\includegraphics[width=0.9\textwidth]{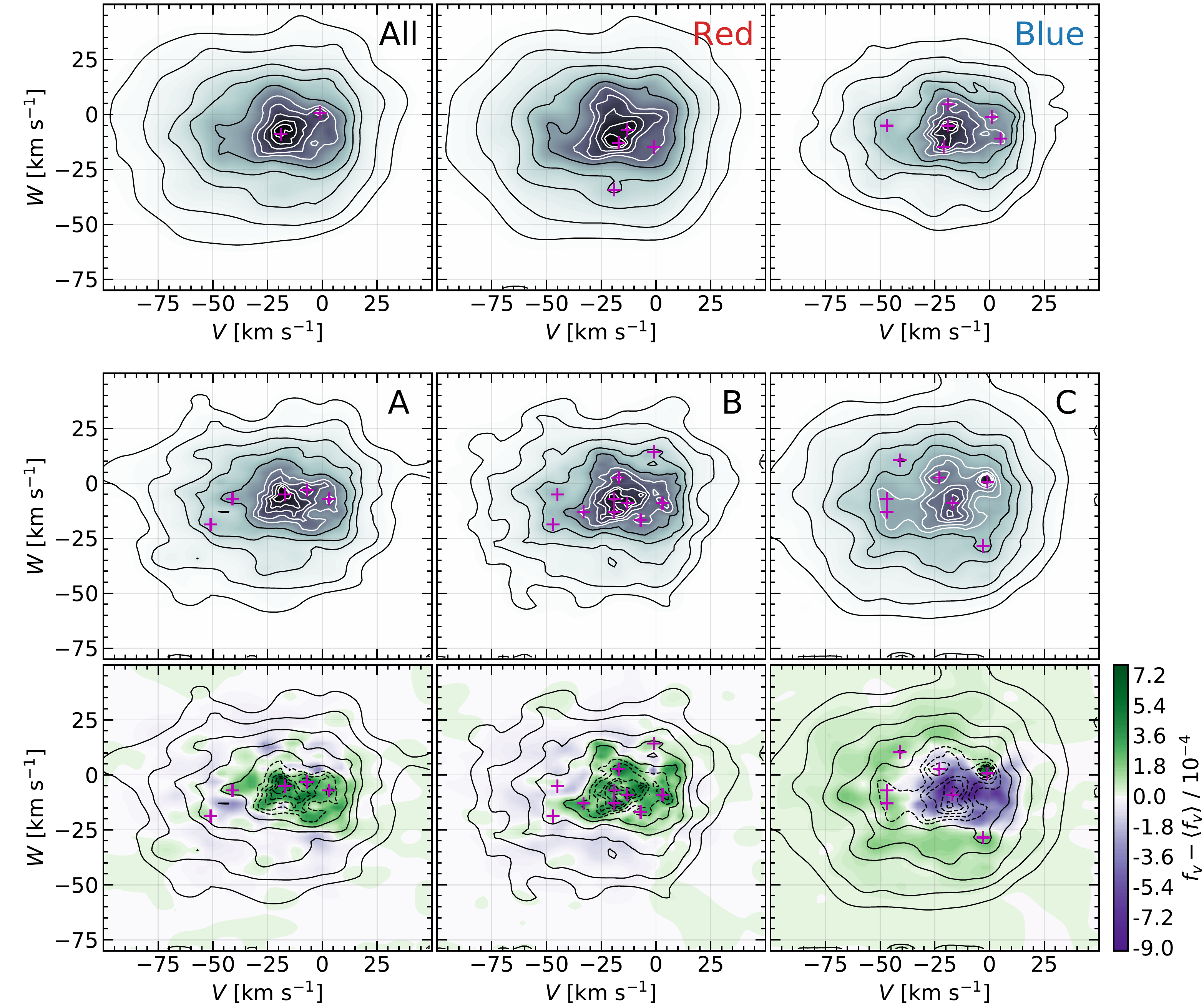}
	\caption{Same as Fig. \ref{fig:VW} using a statistical resample of the proper motions and parallaxes.}
	\label{fig:resample_VW}
\end{figure*}

\bsp	
\label{lastpage}
\end{document}